# Can Bioinformatics Be Considered as an Experimental Biological Science?

Olaf Ilzins, Raul Isea*, Johan Hoebeke

Institute of Advanced Studies (IDEA), Hoyo de la Puerta, Baruta, Venezuela

**Email address**

raul.isea@gmail.com (R. Isea)

**Abstract**

The objective of this short report is to reconsider the subject of bioinformatics as just being a tool of experimental biological science. To do that, we introduce three examples to show how bioinformatics could be considered as an experimental science. These examples show how the development of theoretical biological models generates experimentally verifiable computer hypotheses, which necessarily must be validated by experiments *in vitro* or *in vivo*.

**Keywords**

Bioinformatics, Hypothesis, Experimental, Biological, Computer

## 1. Introduction

What do we mean by "Bioinformatics"?

According to the Collins' dictionary definition, Bioinformatics is defined as "*the branch of information science concerned with large databases of biochemical or pharmaceutical information.*" Although great advances have recently occurred in information technology, the term Bioinformatics is quite often described as just being the process of inserting and storing biological information in a database. This parochial definition limits the term Bioinformatics and possibly condemns it to oblivion rather than indicating an evolution from the definition proposed by Paulien Hogeweg in 1978 [1], modified by Luscombe et al [2], and later remodified by Huerta [3].

Bioinformatics is a science in which the holistic way to understand biology is to intermingle health sciences with information technology without limits or an addressed definition. This means that it is not possible to differentiate the *omic* analysis with respect to the information technology that enables it as is evidenced by a growing body of journals on the subject.

More formally, we suggest that Bioinformatics is conceptualizing biology in terms of interactions too numerically taxing or complex to be organized or analyzed without applying information techniques (themselves derived from applied mathematics, computer science, and statistics). It thus allows the elucidation of relevant information associated with these interactions, furthering our understanding of their relationships and our general scientific knowledge.

Therefore a numerical analysis, mathematical transformations and extrapolations using a scientifically systematic approach based on observed phenomena in or between macromolecules, other interacting agents, or populations, can reveal biologically relevant information. The deduced or inferred information can be submitted to experimental verification. Bioinformatics are thus in themselves scientific experiments since they generate hypotheses that can be experimentally validated or falsified.

## 2. Main Content

Let us introduce three examples to show why Bioinformatics could be considered as an experimental science without forgetting that experimental evidence is necessary to support these hypotheses.

Firstly let's take a look at the evolution from a traditional tool into the gateway to new experimental knowledge. Remember that the first challenge in Bioinformatics was the PCR primer design. Programs that coded a set of observed phenomena on oligonucleotide behavior were developed (for example [4]-[6]). These commonplace programs predicted non-observed sequences that allow an amplification under isothermal conditions. Known as the Loop-mediated isothermal amplification (LAMP) PCR [7], this is a rapid and highly sensitive technique with the advantage of not requiring the extraction of DNA or sophisticated equipment such as a thermocycler. Some examples are useful to confirm a diagnosis in visceral leishmaniasis [8], malaria [9], and so on. This example shows how new knowledge is obtained based on *a priori* Bioinformatics knowledge.

Secondly let us discuss *in silico* inferences that are accepted as the most probable cause in the diffusion process of genetic material through a virus capsid plant, Cowpea Chlorotic Mottle Virus - CCMV. This is a complicated dynamic process studied by snapshots that were obtained from X-ray diffraction data. The nuclear magnetic resonance (NMR) used to register the dynamics of atoms is nearly impossible in view of the tremendous amount of necessitated data processing. Isea et al [10] hypothesized that there was a spontaneous release of genetic material through the CCMV capsid without any energy input and that the infectious mechanism arising from genetic material is thus spontaneous. These results should be validated with experiments *in vivo* similar to the recent work about the prediction in human receptor specificity of H5N1 Influenza A Viruses [11], were the authors developed a phylogenetic algorithm to identify candidate pandemic influenza viruses.

Finally we consider data-driven modeling. Data-driven modeling departs from traditional physical or mathematical modeling in which sets of point observations and time series data are analyzed through one or more different possible approaches drawn from the field of computational intelligence (a sub-branch of Artificial Intelligence that studies adaptive mechanisms to enable or facilitate intelligent behavior in complex and changing environments [12]). The empirical models obtained will be refined by inferring the parameters from the data through iteration cycles that can include new found data, derived from laboratory experiments that test the model based on its own output.

This way to infer and validate multidimensional relationships and interactions is expected to have a great impact in multiple areas in the generation of knowledge. This knowledge is useful for the integration of metabolic pathways in mammalian cells such as hepatocytes [13], for proposing new targets for drug design against Toxoplasma gondii [14] and other parasites, and for gaining insight into the pathophysiology of neoplasic disorders as acute myeloid leukemia [15].

## 3. Conclusion

Bioinformatics could be considered an experimental interdisciplinary science that helps to understand, to obtain and to evaluate new knowledge in biosciences by making use of Information Technology and Computer Science.

Bioinformatics began as a tool to just manage experimental data and perform elementary operations such as identifying any potential similarity between sequences that expanded our knowledge of the biological world. Nevertheless the large volume of information currently available and the advances in computing technology has made it possible to analyze and derive working hypotheses that may allow the scientist to elucidate new biological processes such as the invasion of viral genetic material into the genome.

Current limitations in Bioinformatics can be associated more with those of human knowledge of biological problems than of physical limitations of the computing equipment. On the other hand, biological *in vitro* and *in vivo* experiments remain necessary because bioinformatics cannot as yet replace them, but the conclusions obtained can be a guide to overcome the limitations in human knowledge of biological process.

Bioinformatics by itself cannot compensate for the lack of knowledge of biological processes such as knowing which factors lead to a certain allergies or the development of an effective vaccine against HIV/AIDS or other reemerging diseases such as dengue, malaria, and others. But it enables us to formulate experimentally verifiable hypotheses that may help to understand and elucidate these biological processes and to guide the elaboration of conclusive experiments.

A final reflection regarding the convergence of biosciences and computer science should be made in the application of biological discoveries to computer sciences. Supercomputer centers and farms were born from the supposed necessity of centralized computing power. Nevertheless this kind of implementation is extremely expensive in terms of both money and infrastructure. This computing paradigm is now being toppled by the distributed connectivity networks and patterns very similar to the flow of information in biological systems.

## Acknowledgements

The authors wish to thank Prof. Karl E. Lonngren for his comments the paper.

## References

[1] Hogeweg P. The Roots of Bioinformatics in Theoretical Biology. *PLoS Comput Biol*. 2011; 7(3): e1002021.

[2] Luscombe NM, Greenbaum D, Gerstein M. What is bioinformatics? A proposed definition and overview of the field. *Methods Inf Med*. 2001; 40(4):346-58.

[3] Huerta, M. 2000. NIH working definition of bioinformatics and computational biology. Available from http://www.bisti.nih.gov/docs/compubiodef.pdf [cited: 1 Dec 2014].

[4] Offerman JD, Rychlik W. Oligo Primer Analysis Software in Introduction to bioinformatics: a theoretical and practical approach. Ed. Stephen A. Krawetz and David D. Womble; Humana Press Inc., Totowa, NJ. (2003), pp. 345-361.

[5] Rychlik W, Rhoads RE. A Computer Program for Choosing Optimal Oligonucleotides for Filter Hybridization, Sequencing and in vitro Amplification of DNA. *Nucleic Acids Res* 1989; 17:8543-51.

[6] Rychlik W, Spencer WJ, and Rhoads RE. Optimization of the Annealing Temperature for DNA Amplification in Vitro. *Nucleic Acids Res*.1990; 18:6409-12.

[7]  Notomi T, Okayama H, et al. Loop-mediated isothermal amplification of DNA. *Nucleic Acids Res*. 2000; 28: E63.

[8]  Ghasmian M, Gharavi MJ, Akhlaghi L, Moheball M, Meamar AR, Aryan E and Oormazdi H. Development and Assessment of Loop-Mediated Isothermal Amplification (LAMP) Assay for the Diagnosis of Human Visceral Leishmaniasis in Iran. *Iran J Parasitol*. 2014; 9(1): 50–9.

[9]  Cook J, Aydin-Schmidt B, González IJ, Bell D, Edlund E, Nassor MH, et al. Loop-mediated isothermal amplification (LAMP) for point-of-care detection of asymptomatic low-density malaria parasite carriers in Zanzibar. *Malaria Journal*. 2015; 14: 43.

[10] Isea R, Aponte C, Cipriani R. Can the RNA of the Cowpea Chlorotic Mottle Virus be released through a channel by means of free diffusion? A Test in silico". *Biophysical Chemistry*. 2004; 107 (1-2): 101-6.

[11] Schmier S, Mostafa A, Haarmann T, Bannert N, Ziebuhr J, Veljkovic V, et al. In Silico Prediction and Experimental Confirmation of HA Residues Conferring Enhanced Human Receptor Specificity of H5N1 Influenza A Viruses. *Sci Rep*. 2015; 5: 11434.

[12] Engelbrecht, Andries P. (2007) Computational Intelligence: An Introduction, 2nd Edition. John Wiley & Sons, Ltd.

[13] Huard J, Mueller S, et al. An integrative model links multiple inputs and signaling pathways to the onset of DNA synthesis in hepatocytes. *FEBS Journal* 2012; 279: 3290–313

[14] Tymoshenko S, Oppenheim RD, Agren R, Nielsen J, Soldati-Favre D, Hatzimanikatis V. Metabolic Needs and Capabilities of Toxoplasma gondii through Combined Computational and Experimental Analysis. *PLoS Comput Biol* 2015; 11(5): e1004261.

[15] Levine JH, Simonds EF, et al. Data-Driven Phenotypic Dissection of AML Reveals Progenitor-like Cells that Correlate with Prognosis. *Cell* 2015; 184–97.